\documentclass[twocolumn,pra,preprintnumbers,amsmath,amssymb,superscriptaddress,floatfix]{revtex4}

\usepackage{latexsym}
\usepackage{epsfig}
\usepackage{graphicx}
\usepackage{textcomp}
\usepackage{color}
\usepackage{blindtext}
\usepackage{mathtools}
\usepackage{lipsum}

\setcitestyle{super}
\begin{document}

\title{Optimized communication strategies with binary coherent states over phase noise channels}

\author{M. T. DiMario}
\affiliation{Center for Quantum Information and Control, Department of Physics and Astronomy, University of New Mexico,
Albuquerque, New Mexico 87131}

\author{L. Kunz}
\affiliation{Centre for Quantum Optical Technologies, Centre of New Technologies, University of Warsaw, Banacha 2c, 02-097 Warszawa, Poland}
\affiliation{Faculty of Physics, University of Warsaw, Pasteura 5, 02-093 Warszawa, Poland}

\author{K. Banaszek}
\affiliation{Centre for Quantum Optical Technologies, Centre of New Technologies, University of Warsaw, Banacha 2c, 02-097 Warszawa, Poland}
\affiliation{Faculty of Physics, University of Warsaw, Pasteura 5, 02-093 Warszawa, Poland}

\author{F. E. Becerra}
\affiliation{Center for Quantum Information and Control, Department of Physics and Astronomy, University of New Mexico,
Albuquerque, New Mexico 87131}
\email{fbecerra@unm.edu}

\begin{abstract}
The achievable rate of information transfer in optical communications is determined by the physical properties of the communication channel, such as the intrinsic channel noise. Bosonic phase-noise channels, a class of non-Gaussian channels, have emerged as a relevant noise model in quantum information and optical communication. However, while the fundamental limits for communication over Gaussian channels have been extensively studied, the properties of communication over Bosonic phase-noise channels are not well understood. Here we propose and demonstrate experimentally the concept of optimized communication strategies for communication over phase-noise channels to enhance information transfer beyond what is possible with conventional methods of modulation and detection. Two key ingredients are generalized constellations of coherent states that interpolate between standard on-off keying and binary phase shift keying formats, and non-Gaussian measurements based on photon number resolving detection of the coherently displaced signal. For a given power constraint and channel noise strength, these novel strategies rely on joint optimization of the input alphabet and the measurement to provide enhanced communication capability over a non-Gaussian channel characterized in terms of the error rate as well as mutual information.
\end{abstract}

\maketitle

\noindent
\textbf{INTRODUCTION}
\\
The amount of information that can be transmitted through a physical channel depends on the fundamental properties of the channel \cite{shannon48, holevo73} and the physical states used as information carriers. Recent work has shown that coherent states of light, routinely produced by lasers, can achieve the ultimate limits of information transfer, classical capacity, in communication channels with loss \cite{giovannetti04}, and phase-insensitive Gaussian noise \cite{giovannetti14, mari14}. These results provide strong support for using coherent states as the centerpiece for current and future developments of optical communication networks \cite{he14, kikuchi16,gisin16}. Moreover, beyond the realm of classical communications, coherent states have shown to be of great practical use for quantum communications \cite{arrazola14,arrazola16}, including quantum key distribution \cite{bennett84, bennett92, huttner95, grosshans02, silberhorn02, grosshans03, gisin02, sych10, Takeoka14,pirandola17}, quantum digital signatures \cite{clarke12}, and quantum fingerprinting \cite{xu15b, guan16}.
However, despite the theoretical breakthroughs in identifying the capacities for phase-insensitive Gaussian channels, finding the ultimate information rates for other channels, such as noisy channels with a specific non-Gaussian noise that may be encountered in different situations, is still an open problem. Moreover, even in channels for which capacity is known, reaching this ultimate rate for reliable communications requires finding the optimal encoding schemes and optimal measurements over the physical information carriers \cite{guha11, wilde12}. Furthermore, finding optimal encodings and measurements to maximize information transfer in a specific channel with fundamental noise, in addition to technical noise in real devices, would represent a large advance in our understanding of the limits in realistic optical communications.

Quantum mechanics in principle allows for constructing measurements for coherent states surpassing the classical limits of sensitivity and information transfer \cite{helstrom76, holevo73}. Discrimination strategies for coherent states based on optimized measurements with photon counting have been proposed \cite{dolinar73, kennedy72, taeoka08, bondurant93, becerra11, sych10, nair14, muller15, izumi13} and demonstrated \cite{cook07, wittman08, tsujino11, wittmann10, wittmann10b, muller12, becerra13, becerra15, ferdinand17, dimario18} to surpass the conventional limits of detection, the quantum noise limit (QNL), and approach the ultimate quantum limit, the Helstrom bound \cite{helstrom76}. These nonconventional measurements can enhance information transfer in optical communications \cite{ferdinand17, lee16} and surpass the classical limits of information transfer using joint measurements over sequences of coherent states \cite{guha11}. Furthermore, photon counting measurements can be optimized to provide inherent robustness against noise and imperfections of realistic systems in communications \cite{becerra15, dimario18b}. While these optimized measurements can enhance sensitivities and information transfer with coherent states, the fundamental noise intrinsic in the channel can severely degrade the information encoded in these states. This in turn compromises the potential benefits of these optimized measurements for optical communications \cite{teklu15, jarzyna16}.

In this work, we investigate a new approach for optimizing communications in a channel with specific intrinsic noise in addition to unavoidable technical noise, with the goal of maximizing sensitivities and information transfer based on non-Gaussian measurements and coherent states. The central concept of this approach consists of finding optimized communication strategies where measurements and coherent state encodings are jointly optimized to become more robust to the specific noise in the channel, and ultimately maximize sensitivities and information transfer over the noisy channel. As a proof-of-concept demonstration, we investigate optimized communication strategies for communications over a noisy channel with phase diffusion \cite{genoni11, genoni12, trapani15}, based on optimized single-shot photon-counting measurements and binary coherent state encodings. Phase diffusion is the most detrimental noise for states of light carrying information in the phase, since it destroys the coherence of the quantum states \cite{bina17, chesi18, olivares13}. We show that an optimized strategy that simultaneously optimizes the non-Gaussian measurement and the binary state alphabet allows for surpassing the limits in performance of an ideal conventional measurement in terms of probability of error and information transfer per channel use over the non-Gaussian channel.
\\
\\
\noindent
\textbf{RESULTS}
\\
\textbf{Optimized strategy for a phase diffusion channel}\\
Phase diffusion noise has been extensively investigated in quantum metrology, measurements and communications for phase estimation \cite{genoni11}, interferometry \cite{genoni12}, state discrimination, and information transfer in communication \cite{lau07, ip08, hager13}. This noise is most damaging when information is contained in the coherent properties of the states used as information carriers.
In particular, Gaussian phase diffusion makes the task of extracting information more difficult \cite{genoni11, genoni12, trapani15, bina17, chesi18, olivares13, teklu15, jarzyna14, jarzyna16}, degrading measurement sensitivities and lowering the achievable information transfer in coherent communications.
As a first step for constructing an optimized communication strategy with binary encoding over a channel with phase diffusion, we consider the optimization of the input alphabet to provide robustness to phase diffusion and to other sources of noise and imperfections. This optimization consists of finding the optimal energy distribution in the alphabet to minimize the detrimental effects of phase diffusion, while allowing for measurements to provide high sensitivity.

\begin{figure}[!t]
\centering\includegraphics[width = 8.5cm]{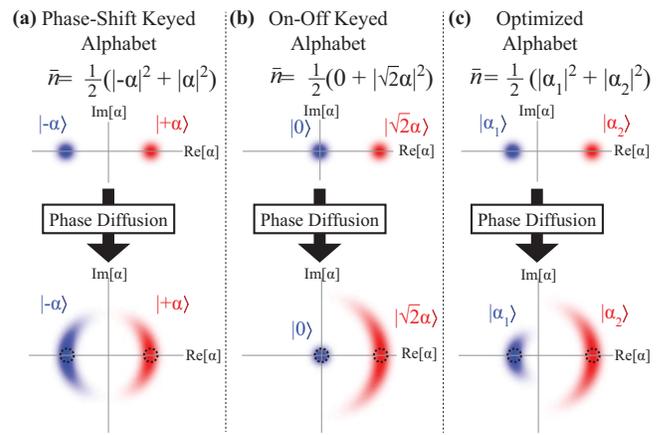}
\caption{\textbf{Phase diffusion for binary alphabets.} (a) Binary phase-shift keyed (BPSK) states, each with mean photon number $\langle n \rangle=\bar{n}$, undergo phase diffusion which equally affects both states as shown in the phase space diagrams. The overlap, and therefore the measurement error, increases with higher noise levels. (b) On-off keyed (OOK) states with total average mean photon number $\bar{n}$. For no noise, the overlap for OOK is greater than BPSK, but it remains constant as phase noise increases. (c) An optimized alphabet such that the total average mean photon number is $\bar{n} = \frac{1}{2}(|\alpha_{1}|^{2} + |\alpha_{2}|^{2})$. Under phase diffusion, one state is affected more than the other. Optimization over the alphabet allows the communication strategy to combine the high sensitivity of BPSK with the robustness to phase noise of an OOK alphabet.}
\label{PhaseDiffExample}
\end{figure}

\begin{figure}[!t]
\centering\includegraphics[width = 8.5cm]{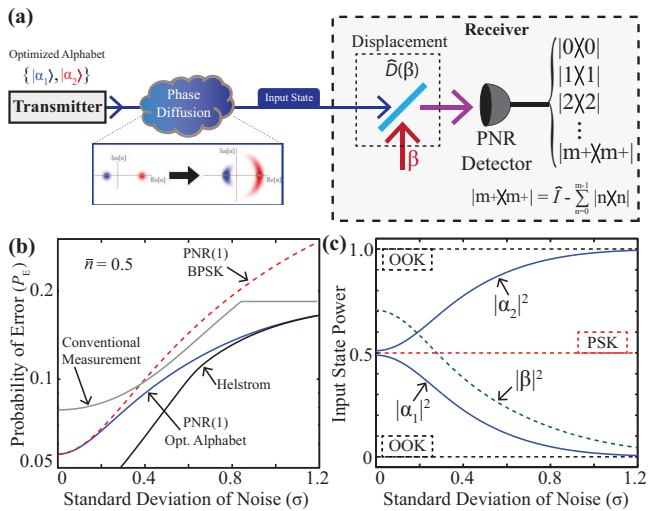}
\caption{\textbf{Optimized communication strategy.} (a) Optimized communication strategy for a channel with phase diffusion with binary coherent state encoding. The receiver uses a single-shot measurement based on optimized photon-number resolving (PNR) detection with finite photon number resolution $m$ (PNR($m$)). For a given level of phase diffusion $\sigma$, the strategy simultaneously optimizes the transmitter's alphabet $\{|\alpha_{1}\rangle,|\alpha_{2}\rangle\}$ and the receiver's discrimination measurement to enhance sensitivities and information transfer through the noisy phase-diffusion channel. (b) Optimized strategy for state discrimination to minimize the probability of error ($P_{\texttt{E}}$) with PNR(1) for an input alphabet with average power $\bar{n} = 0.5$. Probability of error for the optimized strategy (solid blue); for a strategy without input alphabet optimization using BPSK (dashed red); a conventional measurement (CM) with its own optimized alphabet (solid grey); and for the Helstrom measurement with an optimal input alphabet (solid black). (c) Optimized alphabet for the optimized communication strategy (solid blue) and displacement (dashed green) strategy. Note that the optimized alphabet interpolates from BPSK to OOK as the level of phase diffusion $\sigma$ increases. Parameters for the plots: ideal detection efficiency, no dark counts, and an interference visibility of $\xi = 0.998$. }
\label{StrategyFigure}
\end{figure}

Figure \ref{PhaseDiffExample} shows the effect of phase diffusion on three different binary alphabets with coherent states with the same average energy $\langle n \rangle=\bar{n}$: (a) binary phase-shift keying (BPSK) $\{|-\alpha\rangle, |+\alpha\rangle\}$, with $\alpha$ real and positive; (b) on-off keyed (OOK) alphabet $\{|0\rangle, |\sqrt{2}\alpha\rangle\}$; and (c) a general binary coherent state alphabet $\{|\alpha_{1}\rangle, |\alpha_{2}\rangle\}$. We observe that phase diffusion affects equally the states $\{|-\alpha\rangle, |+\alpha\rangle\}$ in the BPSK alphabet, and dramatically reduces their distinguishability, which causes discrimination errors to become very high. On the other hand, when considering the OOK alphabet $\{|0\rangle, |\sqrt{2}\alpha\rangle\}$, phase diffusion impacts only the state $|\sqrt{2}\alpha\rangle$, and leaves the vacuum state $|0\rangle$ unaffected. In this case, their distinguishability weakly depends on the phase noise, highlighting the robustness of this alphabet to phase diffusion noise. Therefore, while BPSK has a smaller overlap and better distinguishability than OOK encoding in the absence of phase noise, OOK states have an overlap independent of the level of phase diffusion.
The optimized alphabet $\{|\alpha_{1}\rangle, |\alpha_{2}\rangle\}$ in Fig. \ref{PhaseDiffExample}(c) represents a smooth transition and a tradeoff between BPSK with a high degree of distinguishability for low levels of noise, and OOK which is immune to phase diffusion. Figure \ref{PhaseDiffExample}(c) shows an example of an optimized alphabet $\{|\alpha_{1}\rangle, |\alpha_{2}\rangle\}$ which is optimized under the average energy constraint $\bar{n} = \frac{1}{2}(|\alpha_{1}|^{2} + |\alpha_{2}|^{2})$ for a given level of the phase noise. The result of this optimization is an alphabet that combines the robustness of OOK with the distinguishability of BPSK.

Optimized non-Gaussian measurements based on photon number resolution (PNR) \cite{dimario18b} provide robustness against technical noise and imperfections for the discrimination of BPSK states surpassing the QNL. Optimized communication strategies in a non-Gaussian channel with phase diffusion can combine these measurements with an optimized input alphabet in order to minimize the probability of error in the channel. This strategy then optimizes simultaneously the measurement and the alphabet, resulting in a high degree of robustness to phase diffusion while maintaining the benefits of non-Gaussian measurements for surpassing the limits of conventional measurements.

Figure \ref{StrategyFigure}(a) shows the concept of an optimized communication strategy for a binary channel with phase diffusion. The sender (Alice) prepares an input state from a coherent state alphabet $\{|\alpha_{1}\rangle, |\alpha_{2}\rangle\}$, and sends it to the receiver (Bob) though a non-Gaussian noisy channel. Phase diffusion causes the input states $\{|\alpha_{k}\rangle\}$ $(k=1,2)$ to become phase diffused mixed states \cite{olivares13}:

\begin{equation}
\label{DiffState}
\hat{\rho}_{k} (\sigma) = \int \displaylimits_{-\infty}^{\infty} \frac{e^{-\frac{\phi^{2}}{2\sigma^{2}}}}{\sqrt{2\pi\sigma^{2}}} | \alpha_{k}e^{-i\phi} \rangle \langle \alpha_{k}e^{-i\phi}| d\phi
\end{equation}
where the strength of the phase diffusion noise is quantified by the width $\sigma$ of the Gaussian phase distribution.

At the channel output, the receiver implements an optimized single-shot measurement based on photon counting to discriminate these states with high sensitivity \cite{dimario18b}. In this strategy, the input state $\hat{\rho}_{k}$ is displaced in phase space to $\hat{D}(\beta)\hat{\rho}_{k} (\sigma) \hat{D}^{\dagger}(\beta)$, where the displacement operation $\hat{D}(\beta) = e^{\beta \hat{a}^\dagger - \beta^* \hat{a}}$ with $\hat{a}$ ($\hat{a}^\dagger$) as the lowering (raising) operator, is implemented by interference of the input state with a displacement field $\beta$ in a high transmittance beam splitter \cite{paris96}. Subsequently, the photons in the displaced state are detected by a photon-number-resolving (PNR) detector with photon number resolution PNR($m$). Here, $m$ represents the maximum number of photons that a detector can resolve before becoming a threshold detector \cite{dimario18b}. This measurement strategy uses a maximum $a$ $posteriori$ (MAP) decision rule to infer the input state based on the photon detection outcome $k$ given the mean photon number $\bar{n}$, displacement field $|\beta|$, and photon number resolution PNR($m$), for a level of phase noise $\sigma$.

The MAP strategy assumes that the correct state is the one with the highest conditional posterior probability $P(\hat{\rho}_{1, 2} (\sigma) |\beta, k, m)$ obtained through Bayes' rule:

\begin{align}
P(\hat{\rho}_{1, 2} (\sigma) |\beta, &k, m) = \frac{P(k |\hat{\rho}_{1,2}(\sigma), \beta, m)P(\hat{\rho}_{1, 2} (\sigma))}{P(k|m)}.
\end{align}

Here, $P(k|m)$ is the total probability of detecting $k$ photons given a PNR($m$) strategy, and $P(k |\hat{\rho}_{1, 2} (\sigma), \beta, m)$ is the conditional probability of detecting $k$ photons given $|\beta|$ and $m$. We consider equiprobable input states, so that the prior probabilities become $P(\hat{\rho}_{1, 2} (\sigma)) = 0.5$. The probability of error in the discrimination of the input states for a strategy with PNR($m$) is:

\begin{align} \label{Pe}
P_\mathrm{E}(\bar{n}, \{\hat{\rho}_{i}(\sigma)\},\beta, m) = \nonumber
\nonumber
\\
1-\frac{1}{2}\sum_{k=0}^{m} \max_{i}(\{&P(k|\hat{\rho}_{i}(\sigma), \beta,m)\}).
\end{align}

Here, $P(k|\hat{\rho}_{i}(\sigma), \beta, \sigma, m)$ is the conditional probability of detecting $k$ photons for the input state given the displacement $|\beta|$, noise level $\sigma$, and PNR($m$). The error probability $P_\mathrm{E}$ in Eq. (\ref{Pe}) depends on the input alphabet, the intrinsic properties of the channel and the measurement performed by the receiver. This provides a way to find optimized strategies that simultaneously optimize the alphabet and the measurement to minimize the detrimental effects of the channel noise. The optimized strategies use an optimal displacement $\hat{D}(\beta)$ and an optimal input alphabet $\{|\alpha_{1}\rangle, |\alpha_{2}\rangle\}$ for a given input power $\bar{n}$, photon number resolution $m$, and channel noise level $\sigma$ to minimize the probability of error $P_{\mathrm{E}}$.

\begin{figure*}[!tbp]
\centering\includegraphics[width = 0.8\textwidth]{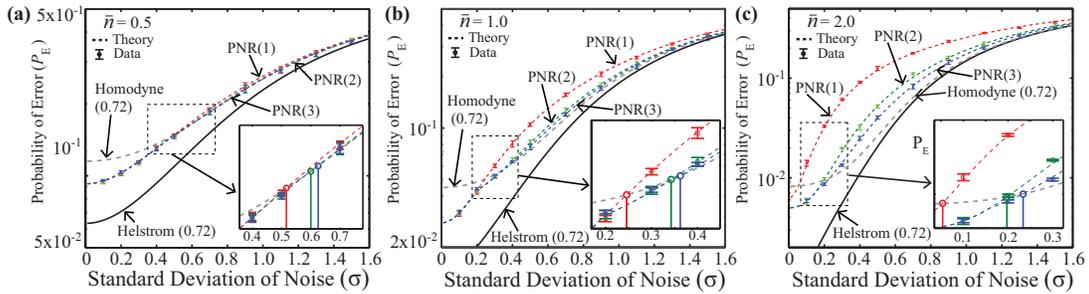}
\caption{\textbf{Experimental error probability for a phase-shift keyed alphabet.} Experimental error probability as a function of standard deviation of the noise $(\sigma)$ for (a) $\bar{n} = 0.5$, (b) $\bar{n} = 1$, and (c) $\bar{n} = 2$, with PNR(1), PNR(2), and PNR(3) (red, green, and blue dots, respectively). Theoretical predictions shown in colored dashed lines. Error bars represent one standard deviation over 5 experimental runs, each consisting of approximately $10^{5}$ independent experiments. Optimized PNR measurements can extend the range of noise for which discrimination below the homodyne limit can be achieved. This benefit becomes larger as the mean photon number increases. The Helstrom bound (solid black line) and the homodyne limit (dashed grey line) are shown adjusted for the system detection efficiency $\eta = 0.72$. Note different vertical scales in figures for different mean photon numbers.
}
\label{dataPEbpsk}
\end{figure*}

Figure \ref{StrategyFigure}(b) shows the performance of an optimized communication strategy for a channel with phase diffusion optimized for state discrimination for a strategy with PNR(1) for $\bar{n} = 0.5$, with ideal detection efficiency $\eta = 1.0$, an interference visibility $\xi = 0.998$ which quantifies the technical noise and imperfections in the receiver \cite{becerra15, dimario18b}, and zero dark count rate $\nu = 0$. To evaluate the performance of this strategy, we compare it with an ideal conventional measurement (CM) consisting of either homodyne or direct detection to minimize the discrimination error, with its own optimized alphabet (solid grey line). We note that the optimized alphabet for the CM results in either BPSK and OOK for this binary coherent state channel, and that it changes abruptly from BPSK to OOK when the conventional measurement switches from homodyne to direct detection.

As shown in Fig. \ref{StrategyFigure}(b), while a PNR(1) strategy with BPSK (dashed red) can only outperform the ideal CM for small phase noise $\sigma$ \cite{olivares13}, optimizing the input alphabet to interpolate between BPSK and OOK, shown in Fig. \ref{StrategyFigure}(c), allows the strategy to outperform the ideal CM for all levels of noise. Moreover, for high levels of noise, the optimized communication strategy approaches the Helstrom measurement with its own optimized alphabet, showing that this optimized communication strategy is asymptotically the optimal quantum measurement.

Optimized communication strategies can also be used to increase information transfer over a noisy channel. These strategies simultaneously optimize the measurement and the input alphabet to maximize mutual information, instead of minimizing probability of error, for a channel with intrinsic noise and technical noise from the devices. Optimized strategies for information transfer for a phase diffusion channel with binary state encoding are in general different from strategies designed for minimum error, as discussed in Section IIIC.
\\
\\
\textbf{Experimental demonstration}\\
The optimized communication strategies described above can be implemented with current technologies. We demonstrate these strategies in a proof-of-principle experiment for enhancing sensitivities and information transfer for the phase diffusion channel with a binary coherent-state encoding with a PNR non-Gaussian measurement, which provides robustness to technical noise and system imperfections \cite{dimario18b}. The experimental realization uses an interferometric setup to implement the optimized strategies. Coherent-state pulses at 633 nm are displaced by interference on a highly transmissive beam splitter, and we use an avalanche photodiode (APD) as a photon number resolving detector. See Ref. \cite{dimario18b} for a detailed description. To investigate the optimized communication strategies, a controlled level of the phase-diffusion noise is applied to the input state (see Supplementary Section 1). Our experiment achieves an overall detection efficiency $\eta = 0.72$, an interference visibility $\xi = 0.998$, and a dark count rate $\nu = 3.6\mathrm{x}10^{-3}$. Technical noise in the experiment such as reduced visibility and dark counts affects the performance of the optimized strategy (see Supplementary Section 2). However, the levels of noise in our experiment only have a small effect on the strategy's performance.

We systematically investigate the optimized communication strategies for a channel with phase diffusion by first studying the performance of optimized PNR measurements with a BPSK alphabet \cite{dimario18b} for this channel. Next we investigate the optimized communication strategies with an optimized measurement-alphabet method for enhancing measurement sensitivity. Finally, we investigate optimized communication strategies for maximizing the mutual information for a phase diffusion channel.
\\
\begin{figure*}[!tbp]
\centering\includegraphics[width = 0.8\textwidth]{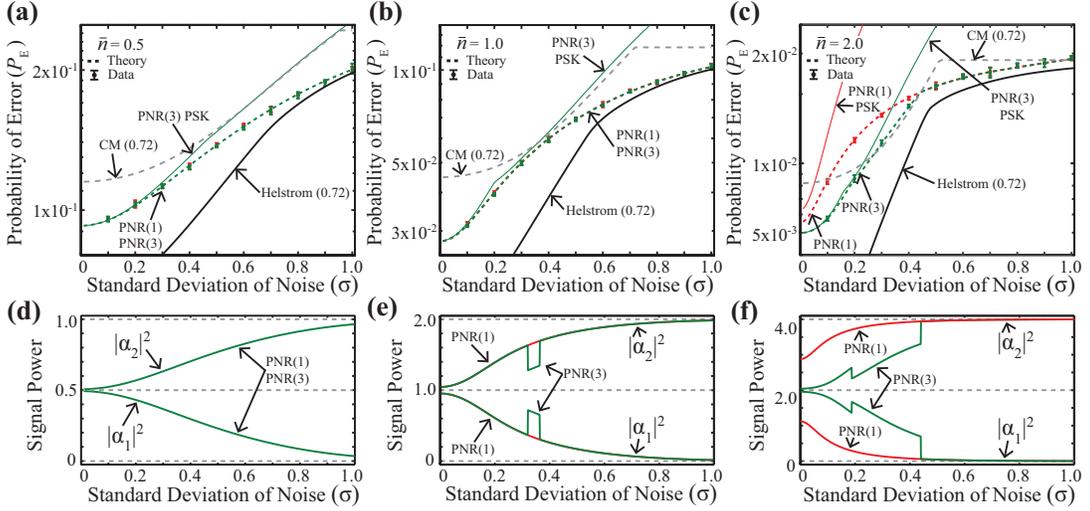}
\caption{\textbf{Experimental error probabilities for an optimized alphabet.} Experimental error probabilities for optimized strategies as a function of standard deviation of the noise $(\sigma)$ for strategies with PNR(1) and PNR(3) (dots) for (a) $\bar{n}  = 0.5$, (b) $\bar{n}  = 1$, and (c) $\bar{n}  = 2$. Included are the Helstrom bound (solid black line) and the ideal conventional measurement (CM) (dashed grey line) both with their own optimized alphabets, and BPSK with PNR(3) measurement (solid green line), all adjusted for our total detection efficiency ($72\%$). While BPSK with PNR(3) can only surpass the CM limit for low levels of noise $\sigma$, optimization of the input alphabet allows for all PNR($m$) strategies to surpass this limit for all noise levels for $\bar{n}  = 0.5$ and $\bar{n}  = 1.0$. (c) For larger $\bar{n}$, increasing the photon number resolution $m$ provides higher robustness for surpassing the CM limit. (d), (e), and (f) show the optimal alphabet for mean photon numbers $\bar{n} = 0.5,~1.0, $ and $2.0$, respectively. }
\label{dataPEopt}
\end{figure*}
\\
\textbf{Discrimination with a BPSK alphabet under phase diffusion}\\
Figure \ref{dataPEbpsk} shows the experimental error probabilities for the discrimination of states from a BPSK alphabet with an optimized PNR measurement \cite{dimario18b} with photon number resolution PNR($m$) of $m=1,2,3$, for three mean photon numbers: (a) $\bar{n} = 0.5$, (b) $\bar{n} = 1$, and (c) $\bar{n} = 2$. We observe in all cases that while PNR(1) (red dots) outperforms an adjusted homodyne measurement up to a certain level of noise, as discussed in Ref. \cite{olivares13}, increasing photon number resolution to PNR(2) (green dots) and PNR(3) (blue dots) extends the level of noise $\sigma$ where this optimized measurement \cite{dimario18b} outperforms a homodyne measurement. The increase in robustness with PNR against phase diffusion becomes larger as the mean photon number increases. Fig. \ref{dataPEbpsk}(b) and (c) show that PNR(3) extends the level of noise $\sigma$ for which this measurement surpasses the homodyne limit by about 1.5 times for $\bar{n} = 1$ and about 4 times for $\bar{n} = 2$ compared to an on/off PNR(1) strategy.
\\
\begin{figure*}[!tbp]
\centering\includegraphics[width = 0.8\textwidth]{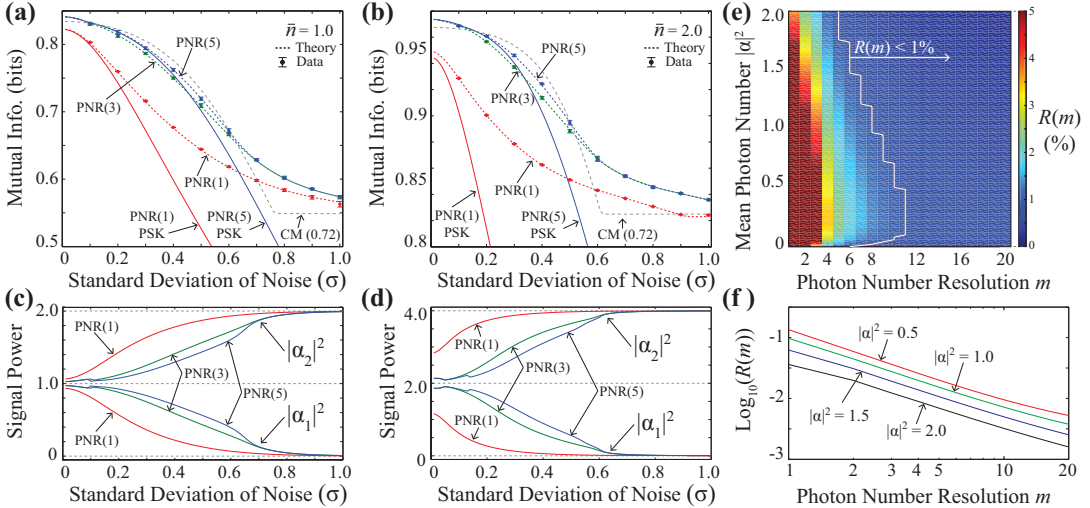}
\caption{\textbf{Experimental mutual information for an optimized alphabet.} Experimentally obtained mutual information $I(X:Y)$ for optimized PNR($m$) strategies with $m$ = 1, 3, and 5 (solid circles) and theoretical predictions (dotted lines) for (a) $\bar{n} = 1.0$ and (b) $\bar{n} = 2.0$. Included are the mutual information for a conventional measurement (CM) and the mutual information with BPSK, all adjusted for our total detection efficiency of $\eta = 0.72$ (dashed grey). Optimized communication strategies with higher number resolution PNR($m$) increase mutual information, and all surpass the CM limit at high levels of phase noise $\sigma$, as well as at low noise $(\sigma \leq 0.1)$. (c,d) Optimal alphabet for (c) $\bar{n}$ = 1.0 and (d) $\bar{n}$ = 2.0. (e) Maximum percent difference of mutual information between the CM and the optimized strategy, $R(m)$, (see main text) with $\eta = 1.0$, $\xi = 0.998$, and $\nu = 0$. The dark blue region to the right of the white line corresponds to $R(m)<1\%$. (f) $R(m)$ on a log-log scale as a function of $m$ for $\bar{n}$ = 0.5, 1.0, 1.5, and 2.0 showing an exponential asymptotic behavior towards $R(m) = 0$ with $m$.}
\label{dataMI}
\end{figure*}
\\
\textbf{Discrimination with an optimized alphabet under phase diffusion}\\
Phase diffusion severely affects measurements for state discrimination in a BPSK alphabet. To reduce the effects of phase diffusion in the channel, a communication strategy can implement an encoding alphabet which is optimized for a particular level of phase noise. In conventional coherent communication with Gaussian measurements, constellation optimization has been used to mitigate some effects of phase noise \cite{lau07, ip08, hager13}. However, in a more general optimized communication strategy using a non-Gaussian measurement, this alphabet can be optimized simultaneously with the displaced photon counting measurement to reduce errors and enhance information transfer.

Figure \ref{dataPEopt} shows the performance of the optimized strategy for the discrimination of states from an optimized alphabet with an optimized PNR measurement \cite{dimario18b} with PNR($1$) and  PNR($3$), for mean photon numbers (a) $\bar{n} = 0.5$, (b)  $\bar{n} = 1.0$, and (c) $\bar{n} = 2.0$. Experimental data is shown with red (green) dots for PNR($1$) (PNR($3$)), and expected performance is shown in dotted lines. Error bars represent one standard deviation over 5 experimental runs of over 10$^{5}$ independent experiments. While a strategy with PNR(3) and a BPSK alphabet (solid green) can only outperform a CM for a limited range of noise levels $\sigma$, optimized strategies with optimal alphabets and measurements allow for outperforming the CM over larger ranges of noise $\sigma$. Moreover, optimized strategies with PNR($1$) surpass the CM for all levels of noise for $\bar{n} = 0.5$ and  $\bar{n} = 1.0$. For higher $\bar{n}$, increasing number resolution $m$ is expected to enable discrimination below the CM at any noise level, as can be inferred from the trend in Fig. \ref{dataPEopt}(c).

Figures \ref{dataPEopt}(d), (e), and (f) show the optimal alphabet for $\bar{n} = 0.5,~1.0, $ and $2.0$, respectively. Discrete jumps in the optimized alphabets for different PNR strategies are the results of optimization of Eq. (\ref{Pe}), which requires a global optimization over multiple minima \cite{dimario18b} of $P_{\texttt{E}}$. This optimization searches for the values of $|\alpha_{1}|$ and $|\beta|$ resulting in the global minimum of $P_{\texttt{E}}$ for a given noise level $\sigma$ for a PNR($m$) strategy. There are levels of noise at which a small increase in $\sigma$ causes the former global minimum of $P_{\texttt{E}}$ as a function of $|\alpha_{1}|$ and $|\beta|$ to become a local minimum, and a former local minimum to become the new global minimum (see Supplementary Section 3). These abrupt changes in the global minimum result in the sudden jumps of the optimal alphabet shown in Fig. \ref{dataPEopt}(e) at $\sigma\approx 0.36$ and $\sigma\approx 0.38$, and in Fig. \ref{dataPEopt}(f) at $\sigma\approx 0.20$ and $\sigma\approx 0.42$. We note that the optimized alphabets correspond to interpolations between BPSK and OOK alphabets for all $\bar{n}$, and result in large improvements over BPSK. This shows that strategies with optimized alphabets are essential or surpassing the sensitivity limits of conventional measurements in the channels with phase noise.
\\
\\
\textbf{Mutual information under phase diffusion}\\
Optimized communication strategies can also be designed to maximize information transfer over a non-Gaussian noisy channel, for which optimal encoding and decoding are unknown. An optimized communication strategy which minimizes probability of error will provide some advantage for increasing mutual information. However, in a noisy channel, the measurement and the alphabet can be optimized in order to maximize mutual information $I(X:Y)$ and will yield a different strategy than for minimum error. Mutual information quantifies the total amount of information between transmitter and receiver, and depends on the encoding alphabet and decoding measurement. 
For a displaced photon-counting measurement, $I(X:Y)$ can be expressed according to a ``soft'' decision rule where the number of photons detected is used to infer the input symbol rather than the binary output from a binary decision rule \cite{mondin15}. The mutual information for a channel with phase diffusion with a binary coherent state encoding can be expressed as:

\begin{widetext}

\begin{equation}
I(\bar{n}, \{\hat{\rho}_{i}(\sigma)\}, \beta, m ) = \sum_{k=0}^{m} \sum_{i=1}^{2} P(k|\{\hat{\rho}_{i}(\sigma)\}, \beta, m)
P(\{\hat{\rho}_{i}(\sigma)\})\log_{2}
\bigg[
\frac{P(k|\{\hat{\rho}_{i}(\sigma)\}, \beta, m)}{P(k|m)}\bigg]
\end{equation}
\end{widetext}

where $  P(k|\{\hat{\rho}_{i}(\sigma)\}, \beta, m)$ is the conditional probability of detecting $k$ photons. In an optimized communication strategy over a noisy channel the input alphabet and measurement with PNR($m$) are simultaneously optimized to maximize mutual information $I(\bar{n},\{\hat{\rho}_{i}(\sigma)\}, \beta, m)$ under the average energy constraint for a noise level $\sigma$.

Figure \ref{dataMI} shows the experimental results for the mutual information with optimized strategies for mean photon numbers (a) $\bar{n} = 1.0$  and (b) $\bar{n} = 2.0$, and photon number resolutions PNR($m$) $m=1,3,5$ in red, green, and blue dots, respectively. The theoretical predictions are shown with dashed colored lines. The mutual information for a conventional measurement (dashed grey), and for BPSK are shown adjusted for our total detection efficiency $\eta = 0.72$. Optimized communication strategies surpass the limit in mutual information for a CM at high levels of phase diffusion noise ($\sigma \geq 0.7$), and for low noise $(\sigma \leq 0.1)$. Moreover, optimized strategies with higher PNR detection resolution $m$ provide higher mutual information for all levels of noise. Note that optimized communication strategies with optimized alphabets drastically outperform BPSK for all PNR ($m$) in terms of mutual information. Fig. \ref{dataMI}(c,d) show the optimized alphabets for (c) $\bar{n} = 1.0$, and (d) $\bar{n} = 2.0$, respectively. We observe that the optimal alphabet interpolates from BPSK to OOK similar to error probability. However this interpolation is continuous, because the mutual information is a convex function of $\sigma$ for all PNR($m$). In the intermediate level of noise $(\sigma \approx0.5)$, there is a gap between the optimized strategies and the CM. This gap decreases as the photon number resolution PNR($m$) of the optimized strategies increases. This suggests that optimized communication strategies with high-enough photon number resolution $m$ should provide levels of mutual information at least as high as those that can be achieved with ideal conventional measurements for all levels of phase diffusion noise.

Figure \ref{dataMI}(e) shows the maximum percent difference $R(m)$ between an optimized strategy with PNR($m$) and a CM for $\bar{n}$ from 0 to 2.0 for different PNR($m$) from $m$=1 to $m$=20. This corresponds to the percent difference at the level of noise for which a PNR($m$) strategy has the worst performance relative to a conventional measurement. $R(m)$ is defined as:

\begin{equation}
\label{Rmin}
R(m) = \max_{\sigma}\left( \frac{I_{CM}(\sigma) - I_{PNR(m)}(\sigma)}{I_{CM}(\sigma)} \right),
\end{equation}

where $I_{PNR(m)}(\sigma)$ is the mutual information for an optimized communication strategy with PNR($m$), and $I_{CM}(\sigma)$ is the mutual information for the conventional measurement. We observe that as the number resolution increases, the percent difference asymptotically approaches zero for all mean photon numbers. The blue regions to the right of the white line correspond to $R(m) < 1\%$, i.e. when a PNR($m$) strategy is within $1\%$ of the conventional measurement. Figure \ref{dataMI}(f) shows $R(m)$ on a log-log scale for $\bar{n}$ = 0.5, 1.0, 1.5, and 2.0 in red, green, blue, and black lines, respectively. The straight lines indicate power-law scaling in the convergence of the form $a(m)^{b}$, with $b \approx 1.1$ for all lines. This convergence suggests that for all mean photon numbers, optimized communication strategies with large enough photon resolution $m$ will at worst provide the same mutual information as the ideal CM, which serves as a lower bound for the performance of optimized communication strategies. At the same time, these optimized strategies with moderate photon number resolution provide large advantages for increasing mutual information compared to CM at low noise and high noise levels.
\\
\\
\noindent
\textbf{DISCUSSION}
\\
We proposed and demonstrated optimized communication strategies to maximize information transfer and measurement sensitivity over a non-Gaussian noisy channel. These optimized strategies are based on simultaneous optimization of the states used as information carriers with an optimized non-Gaussian photon counting measurement that surpasses the QNL for state discrimination. Simultaneous optimization of alphabet and measurement provides robustness to intrinsic channel noise, and allows for overcoming the sensitivity limits of conventional measurements and achieving higher information transfer in communications over noisy channels.
\\
\indent We demonstrated in a proof of principle experiment the concept of optimized strategies for communication over a channel with phase diffusion for binary coherent state alphabets and single-shot optimized measurements with photon number resolution. These optimized communication strategies provide unexpected benefits to minimize the probability of decoding error and maximize the achievable mutual information in this noisy channel. Moreover, we observed that optimized communication strategies not only provide robustness to intrinsic channel noise, but also to technical noise and imperfections in the receiver.
\\
\indent We expect that optimized communication strategies can provide advantages for different problems in coherent communications extending to communication with multiple states and complex measurements. Moreover, optimized communication strategies can be applied to other channels utilizing practical optimized measurements and encodings to maximize information transfer in realistic noisy communication channels for which capacity limits are unknown, but that are encountered in optical communication networks.
\\
\\
\noindent
\textbf{ACKNOWLEDGEMENTS}
\\
This work was supported by the National Science Foundation (NSF) (PHY-1653670, PHY-1521016), and the project ``Quantum Optical Communication Systems'' carried out within the TEAM program of the Foundation for Polish Science co-financed by the European Union under the European Regional Development Fund.
\\
\\
\noindent
\textbf{AUTHOR CONTRIBUTIONS}
\\
F.E.B. and K.B. conceived the idea and supervised the work. L.K. and M.T.D. conducted the theoretical study.
M.T.D. designed the experimental implementation and performed the measurements.
All authors contributed to the analysis of the theoretical and experimental results and contributed to writing the
manuscript.
\\
\\
\noindent
\textbf{COMPETING INTERESTS}
\\
The authors declare that there are no competing interests.
\\
\\
\textbf{DATA AVAILABILITY}
\\
The data that support the findings of this study are available from the authors upon
request.
\\
\\
\noindent
\textbf{REFERENCES}

\newpage

\clearpage

\onecolumngrid

\setcounter{equation}{0} \setcounter{figure}{0}
\setcounter{table}{0} \setcounter{page}{1} \makeatletter
\renewcommand{\theequation}{S\arabic{equation}}
\renewcommand{\thefigure}{S\arabic{figure}}
\renewcommand{\thetable}{S\arabic{table}}

\renewcommand{\bibnumfmt}[1]{[#1]}
\renewcommand{\citenumfont}[1]{#1}

\begin{center}
	
	\textbf{\Large {Supplementary Material}}\\[.2cm]
	
\end{center}

\section{Phase diffusion preparation and calibration}

Gaussian phase noise with controllable amplitude and bandwidth is prepared in the input state using an arbitrary function generator and a phase modulator, which modulates the phase of the input state.
We use the interference of the input field with the local oscillator (LO) field with a given relative phase to estimate the strength of Gaussian phase noise by observing the photon number distributions with an avalanche photodiode.

\begin{figure*}[!bp]
	\centering\includegraphics[width = 0.6\textwidth]{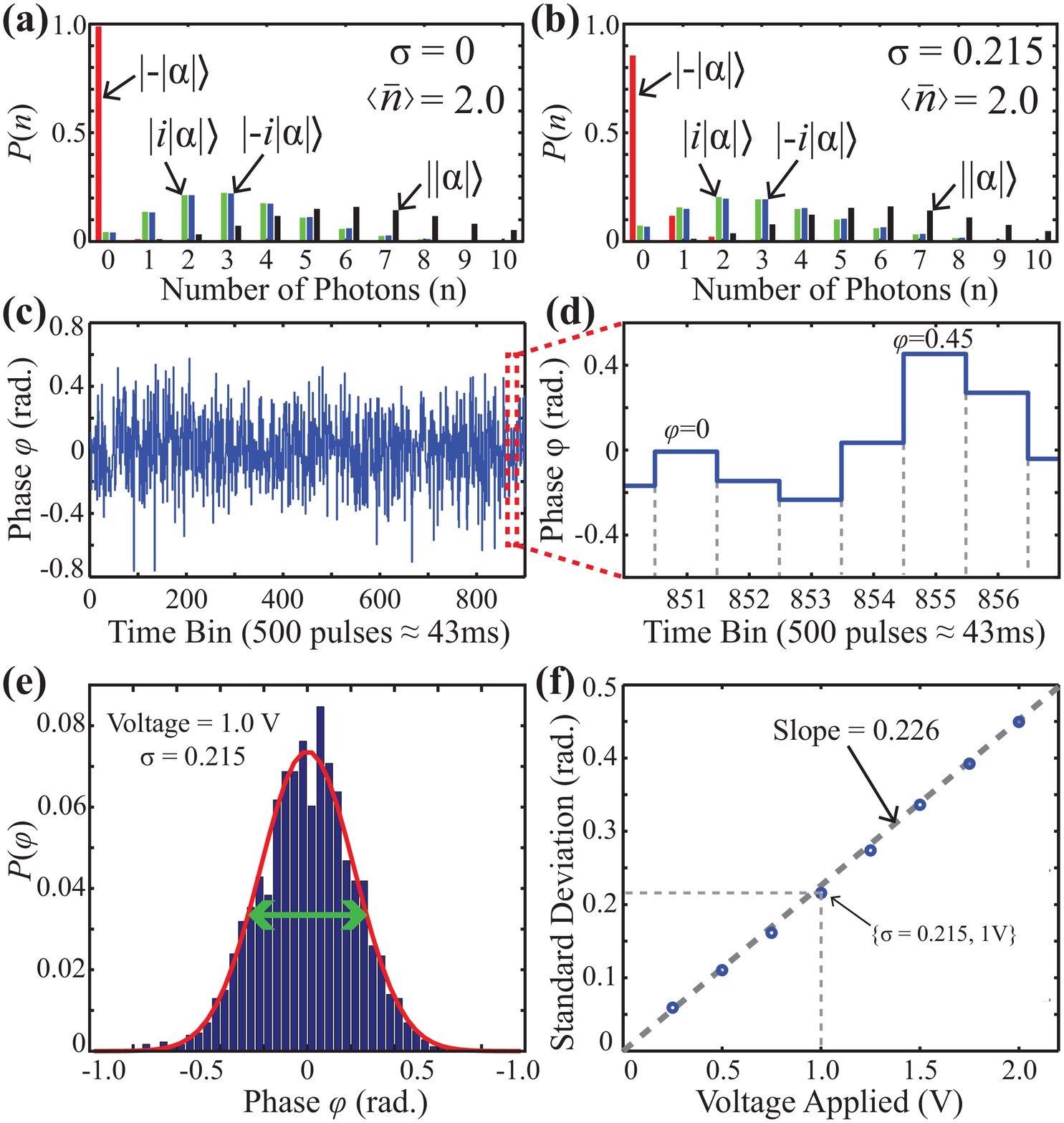}
	\caption{\textbf{Calibration of phase diffusion strength $\sigma$.} Photon number distribution for input states $\{{|-\alpha\rangle}$$, {|i\alpha\rangle}, {|-i\alpha\rangle}, {|\alpha\rangle}\}$ with $\langle n \rangle = 2.0$ in red, green, blue, and black, respectively, displaced by $\hat{D}(\alpha)$, (a) without and (b) with phase diffusion with $\sigma = 0.215$. (c) Time trace of the phase extracted from the photon number distributions with phase noise $\sigma = 0.215$, for the input state and LO with a relative phase of $\phi = \pi/2$. Note that the vertical axis has been shifted so that it shows deviations from $\pi/2$. (d) Zoom in of the trace showing piecewise constant phases over time bins, each containing 500 pulses. (e) Histogram of phases extracted from the time traces, from which the standard deviation $\sigma$ of the phase noise is estimated for a given voltage amplitude applied from the function generator. (f) Plot of the extracted standard deviation of the noise as a function of the applied voltage. The points are fitted to a line, and the slope allows for relating applied voltage levels to phase noise levels precisely.}
	\label{calibration}
\end{figure*}

Figure \ref{calibration}(a) and (b) show examples of the photon number distributions for input states ${|-\alpha\rangle}, {|i\alpha\rangle}, {|-i\alpha\rangle},$ and $|\alpha\rangle$ with $\langle n \rangle = 2.0$, displaced by $\hat{D}(\alpha)$ without (a) and with (b) phase diffusion with $\sigma = 0.215$. Phase diffusion modifies the photon number distribution for different input states, which can be used to estimate the level of induced noise $\sigma$. For example, while the input state $\hat{\rho}_{1}(0)=|-\alpha\rangle\langle-\alpha|$ is ideally displaced to vacuum by $\hat{D}(\alpha)$ when $\sigma=0$, phase diffusion modifies the photon number distribution to show support over higher numbers of detected photons, see Fig. \ref{calibration}(b). The calibration of the phase noise of the input state consists of (1) applying a piecewise constant Gaussian waveform to the phase modulator and estimating the distribution of induced phases, and (2) using a Gaussian fit to estimate the standard deviation $\sigma$ of the phase distribution, which quantifies the level of phase noise.

Figure \ref{calibration}(c-f) shows an example of the calibration of phase diffusion with $\sigma = 0.215$ using states $| \pm i\alpha \rangle$ with mean photon number $\bar{n} = 2.0$. The interference of these states with the LO with phase 0, allows for calibrating the phase noise at relative phases of $\phi = \pi/2$ and $3\pi/2$, which are the points that provide the highest sensitivity. The piecewise constant phase noise applied to the input state allows for defining time bins over which the relative phase of the input state and the LO is constant. Each time bin has a length of $T \approx 43$ ms and contains 500 shots of the experiment all with the same relative phase. For each time bin we measure photon number detections and the relative phase for each time bin is extracted through the mean of the measured photon number distribution during that time bin:
\begin{equation}
\label{pn}
\langle n \rangle_{\pm} = 2\eta \langle n \rangle (1 \pm \xi \mathrm{sin(\hat{\phi})})
\end{equation}
Figure \ref{calibration}(c) shows the reconstructed relative phase as a function of time with piecewise constant Gaussian phase noise for relative phase $\phi = \pi/2$. A zoom in time in Fig. \ref{calibration}(d) shows time bins with constant phases over $T \approx 43$ ms (500 pulses). Note that the vertical axis has been shifted with respect to $\phi = \pi/2$ so that it shows deviations from $\pi/2$. These phases are expected to be Gaussian distributed, which can be used to estimate and calibrate the level of induced Gaussian phase noise.

Figure \ref{calibration}(e) shows the histogram of extracted phase, combined for relative phases $\phi = \pi/2$ and $3\pi/2$, extracted from the photon number distributions with $\langle n \rangle = 2.0$, for a waveform with amplitude of 1 V from the function generator. The fit to a Gaussian distribution results in a standard deviation of $\sigma = 0.215$, which quantifies the level of phase noise for this voltage. We repeat this procedure for different voltage levels of the function generator to calibrate the induced phase noise level as a function of applied voltage. Fig. \ref{calibration}(f) shows the level of phase noise $\sigma$ as a function of applied voltage from the function generator, showing a linear relationship. A fit to a straight line allows for determining the relation between applied voltage and induced phase, which can be used for precise preparation of phase noise of the input state in the experiment.

\begin{figure*}[!thbp]
	\centering\includegraphics[width = 0.65\textwidth]{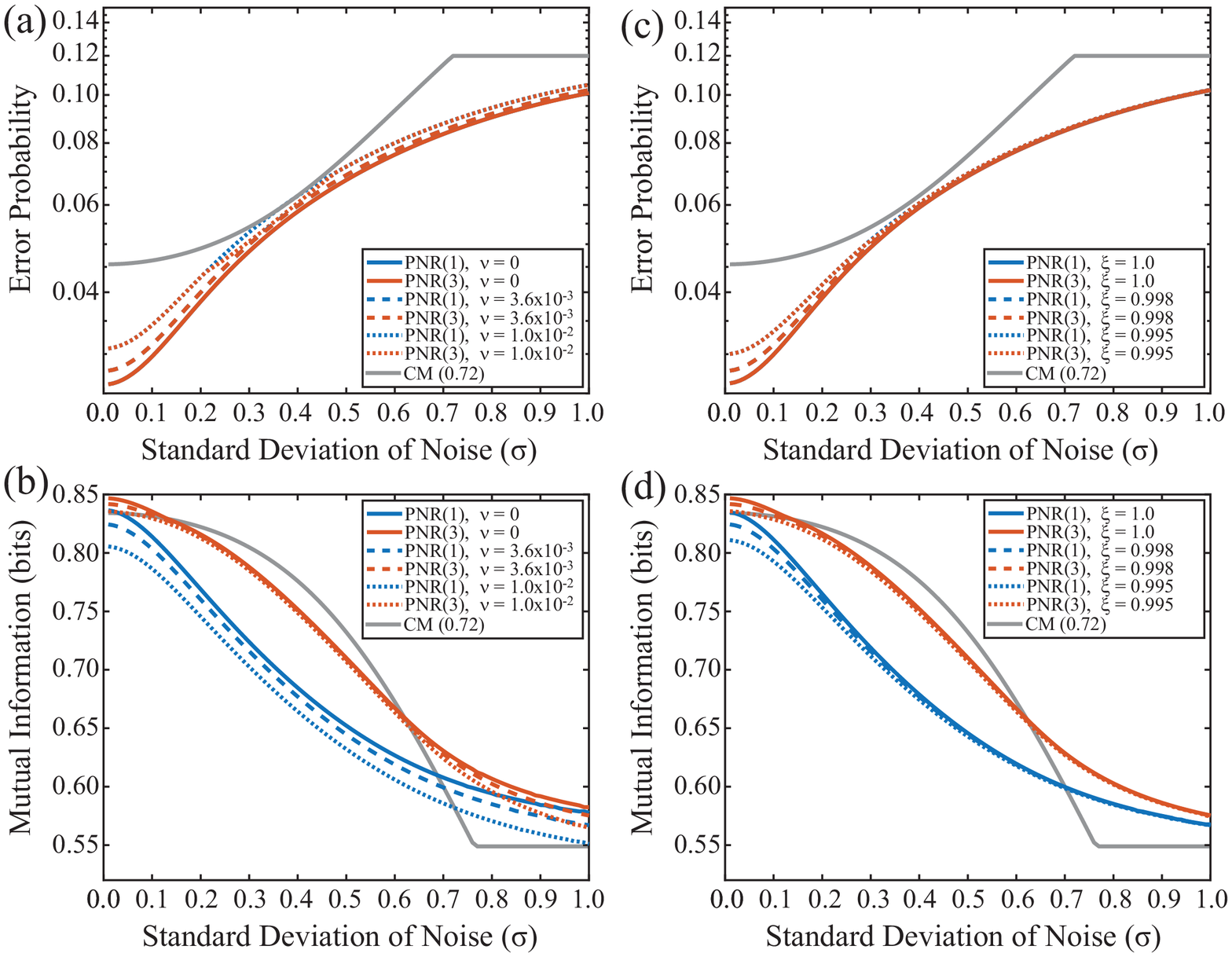}
	\caption{\textbf{Effects of dark counts and reduced visibility.} Optimized communication strategy with detector dark count rates of $\nu =0,~3.6\mathrm{x}10^{-3}$, and $1\mathrm{x}10^{-2}$ for (a) minimizing probability of error, and (b) maximizing mutual information. Optimized communication strategies with reduced visibility for visibilities $\xi =1,~0.998$, and $0.995$ for (c) probability of error and (d) mutual information. The plots are shown for $\bar{n}= 1.0$, which corresponds to an example in our experimental implementation. Note that for levels $\nu = 3.6\mathrm{x}10^{-3}$ and $\xi = 0.998$ in our experiment there is only a slight degradation in the expected performance of the strategies. Here CM is an ideal conventional measurement.}
	\label{noise}
\end{figure*}

\section{Effects of dark counts and reduced visibility}

The performance of a communication strategy depends on the noise and imperfections in any implementation. In our experiment, the main sources of noise and imperfections are detector dark counts and system imperfections resulting in mode mismatch between input state $\hat{\rho}_{k}$ and displacement field $|\beta\rangle$, which can be accounted for by a reduced visibility \cite{becerra15}. Fig. \ref{noise} shows the expected performance of the optimized communication strategies in the phase diffusion channel for an average photon number $\bar{n}= 1.0$ for different levels of detector dark counts ($\nu$) in Fig. \ref{noise}(a) and (b) and reduced visibility ($\xi$) in Fig. \ref{noise}(c) and (d), for probability of error and mutual information, respectively. An average energy of $\bar{n}= 1.0$ corresponds to an example of the input power in our experimental implementation. We observe that higher dark counts and lower visibility degrade the performance of the optimized strategy, increasing the probability of error and reducing mutual information. However, we observe that for the levels of noise and imperfections in our experiment with a dark count rate $\nu = 3.6\mathrm{x}10^{-3}$ and a visibility $\xi = 0.998$, these imperfections only have a small effect on the strategy's performance. We also note that the detector's after pulsing in our experiment ($\approx 1\%$) has a very small effect on the optimized strategy, as has been observed for optimized PNR measurements for state discrimination in a pure-loss channel \cite{dimario18b}.

\begin{figure*}[!tbp]
	\centering\includegraphics[width = 0.7\textwidth]{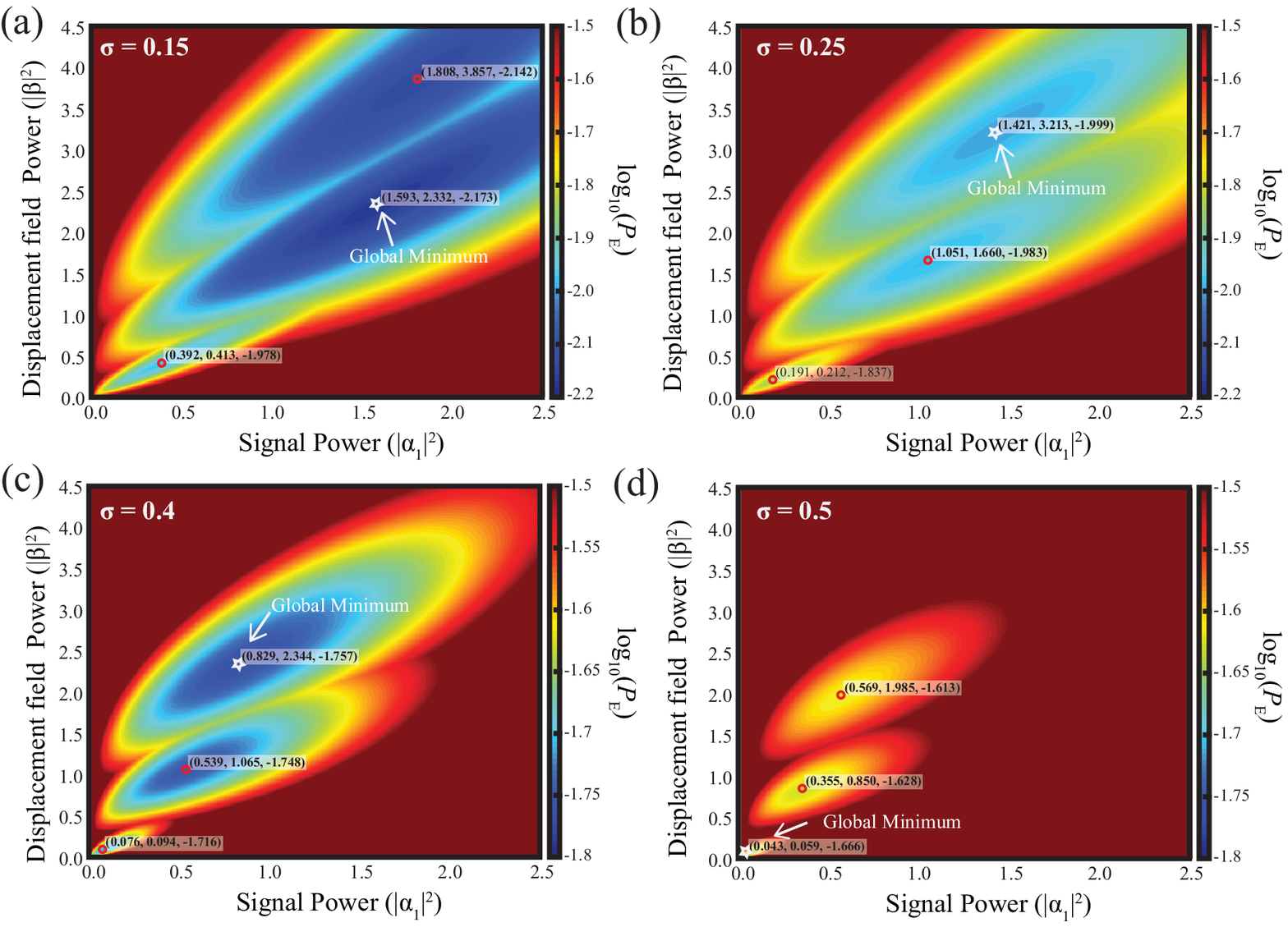}
	\caption{\textbf{Probability of error for optimized strategies.} Probability of error plotted as logarithm ($\log_{10}{P_\mathrm{E}}$) as a function of the power of the input state $|\alpha_{1}|^2$ and displacement field $|\beta|^2$ for PNR(3) and levels of noise (a) $\sigma=0.15$, (b) $\sigma=0.25$, (c) $\sigma=0.4$, and (d) $\sigma=0.5$. The probability of error accounts for dark counts, visibility, and detection efficiency in our experiment to match the situation shown in Fig. 4(d) in the main manuscript, which shows two discrete jumps in $|\alpha_{1}|$: one at $\sigma\approx0.2$, and a second one at $\sigma\approx0.42$. These levels of noise show the change in global minima in $P_\mathrm{E}$ when the noise $\sigma$ increases from $\sigma=0.15$ to $0.25$, and from $\sigma=0.4$ to $0.45$. Labels show values for $\{|\alpha_{1}|^2,~|\beta|^2,~P_\mathrm{E}\}$ for local minima (red circles) and global minima (white stars).
	}
	\label{OptAlphfig}
\end{figure*}

\section{Optimized Alphabet}

The optimal alphabets $\{|\alpha_{k}\rangle\}$ ($k=1,2$) for the optimized strategies are obtained by minimizing the probability of error $P_\mathrm{E}(\bar{n}, \{\hat{\rho}_{i}(\sigma)\},\beta, m)$ in Eq. (3) in the main manuscript. $P_\mathrm{E}$ depends on the input powers $|\alpha_{k}|^2$, the level of noise $\sigma$, the optimized displacement field $|\beta|$, and the photon number resolution $m$ of the detection strategy PNR($m$). $P_\mathrm{E}$ as a function of the input signal power $|\alpha_{1}|^2$ ($|\alpha_{2}|^2=\bar{n}-|\alpha_{1}|^2$) and $|\beta|^2$ for a strategy with number resolution $m$, PNR($m$), is a function with multiple minima, showing a total of $m$ minima. Figure \ref{OptAlphfig} shows an example of the logarithm of the probability of error $\log_{10}(P_\mathrm{E})$ as a function of $|\alpha_{1}|^2$ and $|\beta|^2$ for average photon number $\bar{n}=2$ for PNR($3$), for noise levels (a) $\sigma=0.15$, (b) $\sigma=0.25$, (c) $\sigma=0.4$, and (d) $\sigma=0.45$. Note that for each value of $\sigma$ there are three minima: two local minima (red circles), and one global minimum (white stars). The probability of error in Fig. \ref{OptAlphfig} accounts for the dark counts $\nu=3.6\mathrm{x}10^{-3}$, visibility $\xi=0.998$, and detection efficiency $\eta=0.72$ in our experiment. This situation corresponds to the case shown in Fig. 4(c) and (f) in the main manuscript for $\bar{n}=2$, which shows two discrete jumps of the optimized alphabet $|\alpha_{1}|^2$ at $\sigma\approx0.2$ and at $\sigma\approx0.42$. We observe in Fig. \ref{OptAlphfig} that by increasing the noise $\sigma$ from 0.15 to 0.25, there is a change in which minima is the global minimum. This sudden change causes the optimized alphabet $\{|\alpha_{k}\rangle\}$ to show a discrete jump between these two values, as can be seen in Fig. 4(f) in the main manuscript around $\sigma\approx0.2$. In the same way, by increasing $\sigma$ from 0.4 to 0.5, there is a change in global minima, causing a discrete jump of $\{|\alpha_{k}\rangle\}$ around $\sigma\approx0.42$. Our numerical studies show that for $\bar{n}=1$, we expect two discrete jumps in the optimized alphabet at $\sigma\approx0.36$ and $\sigma\approx0.38$ (see Fig. 4(e) in the main manuscript). However,  for $\bar{n}=0.5$ there is not any change of global minimum of $P_\mathrm{E}$ for PNR(3), so there are not expected discrete jumps in the optimized alphabet $\{|\alpha_{k}\rangle\}$, as can be seen in Fig. 4(d) in the main manuscript.

The optimization of the probability of error is a highly nonlinear function, since it is based on the maximum a-posteriori probability criterion \cite{dimario18b}. This causes discrete jumps in the optimized alphabet for PNR($m$), $m>1$. On the other hand, the mutual information $I(\bar{n}, \{\hat{\rho}_{i}(\sigma)\}, \beta, m )$ in Eq. (4) in the main manuscript is a smooth function of $|\alpha_{k}|^2$ and $|\beta|^2$ with a single maximum. As a result, the optimized alphabets  $\{|\alpha_{k}\rangle\}$ for maximizing mutual information in phase-noise channels do not show any discrete jumps.
\\

\textbf{REFERENCES}

\end{document}